\documentclass[11pt]{article}
\usepackage{moriond,epsfig}

\def\be{\begin{equation}}
\def\ee{\end{equation}}
\def\bea{\begin{eqnarray}}
\def\eea{\end{eqnarray}}

\begin{document}
\vspace*{4cm}
\title{NEUTRINO ASYMMETRY GENERATION IN THE EARLY UNIVERSE 
(FROM $\nu_{\alpha}\leftrightarrow \nu_s$ OSCILLATIONS)}

\author{P. DI BARI }

\address{INFN, Italy and School of Physics, University of Melbourne, Vic 3010, Australia.}
\maketitle
\abstracts{Neutrino asymmetry generation from active-sterile neutrino
oscillations in the early Universe has different important cosmological effects.
We discuss its basic features and present recent results on the borders of the 
generation for mixing angles $\sin^2 2\theta_0\leq 10^{-7}$.}

\section{General conditions for neutrino asymmetry generation}

Let us consider, from a statistical point  of view, 
a two neutrino mixing $\nu_{\alpha}\leftrightarrow\nu_{\beta}$ 
in the early Universe as a special elementary process violating 
flavor lepton number conservation but still preserving the 
total lepton number conservation in a way that 
$L_{\nu_{\alpha}}+L_{\nu_{\beta}}={\rm const}$. 
Since the amplitude of the direct processes is 
equal to that one of the inverse processes, one expects that at the equilibrium 
the number of $\nu_{\alpha}$ is equal to that of $\nu_{\beta}$ and
the same for anti-neutrinos. 
This clearly implies also the equality of the asymmetries
\footnote{We define the $X$ particle species asymmetry $L_X\equiv (N_X-N_{\bar{X}})/N_{\gamma}^{\rm in}$, with
$T_{\rm in}\sim 10\,{\rm MeV}$.}
in a way that $L_{\nu_{\alpha}}^{\rm eq}=L_{\nu_{\beta}}^{\rm eq}={\rm const}/2$.
There is however the possibility that neutrinos and anti-neutrinos 
mixings are different. In this case the two different equilibrium conditions for
particle numbers can be reached at different times for neutrinos and anti-neutrinos. 
In the case that $\alpha$ and $\beta$ are two 
ordinary neutrino flavors, they are both in thermal equilibrium conditions.
This means that if initially $L_{\nu_{\alpha}}=L_{\nu_{\beta}}=0$, then 
also particle numbers are necessarily equal.
This is why an asymmetry cannot be generated from active-active neutrino oscillations.
This is not true if one of the two flavors is a sterile neutrino flavor $\nu_s$ whose number density is negligible compared to that of active neutrinos at temperature below the 
quark-hadron phase transition at around $150\,{\rm MeV}$.  In this way if active neutrinos 
and active antineutrinos are converted into sterile neutrinos and anti-neutrinos at different 
rates, the equilibrium for the particle numbers can be reached at different times
for neutrinos and anti-neutrinos. This implies that during this {\em relaxation phase} 
the neutrino asymmetries can change. In particular if one assumes that at the beginning
$L_{\nu_{\alpha}}=L_{\nu_s}=0$, then, during this approach to equilibrium, one can have
$L_{\nu_{\alpha}}=-L_{\nu_s}\neq 0$: an $\alpha$-neutrino asymmetry has been generated.
This generation would be just a temporary phase if in the end the equilibrium
is reached. But if the process $\nu_{\alpha}\leftrightarrow\nu_s$ 
freezes before, then an $\alpha$ neutrino asymmetry would eventually survive.
Note that the generation of a neutrino asymmetry has to respect the same Sakharov
conditions for baryogenesis, 
namely the existence of a process that violates charge 
(in our case $\alpha$-lepton number) conservation, 
out-of-thermal equilibrium conditions and $CP$ violation, 
that would make possible a different behaviour of neutrino and anti-neutrinos.
An early investigation was done in \cite{Petcov}, but matter effects were neglected
and conclusions were negative. If matter effects are taken into account, 
$CP$ violation is provided for free by the presence of the baryon
asymmetry. This enters the {\em total asymmetry} of the medium, $L^{(\alpha)}$,
in the Wolfenstein term of the effective potential \cite{Notzold} 
but with opposite sign for neutrinos and anti-neutrinos.
This quantity includes also the neutrino asymmetries and in particular the
$\alpha$-neutrino asymmetry itself. It means that if the $\alpha$ neutrino asymmetry
changes, then matter effects change. Therefore there is another condition
that has to be satisfied for the generation: a condition of {\em instability}.
Let us assume, for definiteness, that the total asymmetry of all non oscillating particle species, is positive and let us indicate it with $\tilde{L}$ and that the initial
$\alpha$ neutrino asymmetry is zero in a way that $L^{(\alpha)}_{\rm in}=\tilde{L}$.
If this initial positive total asymmetry makes neutrino mixing slightly stronger 
than the anti-neutrino mixing, a negative $\alpha$-neutrino asymmetry starts to be 
generated but the total asymmetry, 
$L^{(\alpha)}\equiv 2\,L_{\nu_{\alpha}}+\tilde{L}$, is destroyed 
and $L_{\nu_{\alpha}}^{\rm fin}=-\tilde{L}/2$. 
We know that the baryon asymmetry is tiny and thus only assuming that the non oscillating neutrinos asymmetries are much larger one can generate a large (${\cal O}(0.1)$) 
$\alpha$-neutrino asymmetry in this way. Maybe this can have some applications,
but it would be much more interesting if, assuming initial tiny neutrino asymmetries
(for example of the order of the baryon asymmetry) a large $\alpha$ neutrino asymmetry
can be generated. Then one necessarily has to look for a situation in which, given
a (still positive for definiteness) initial total asymmetry 
$L_{\rm in}^{(\alpha)}=\tilde{L}$, 
the anti-neutrino mixing is stronger than neutrino mixing. 
This implies that more active anti-neutrinos are converted
and thus that a positive neutrino asymmetry is generated and this time 
the total asymmetry $L^{(\alpha)}$ also increases.
In the case of positive $\delta m^2$ one has always a stable behaviour.
While for negative $\delta m^2$ it can be shown that when the resonant
(rescaled) momentum of neutrinos $y_{\rm res}^{\nu}\equiv p_{\rm res}^{\nu}/T$
(that increases as $T^{-3}$ and is equal to that of anti-neutrinos until the asymmetry remains small enough) becomes $\sim 2$, then an instability takes place.
 We are therefore interested to negative values of $\delta m^2$
(`sterile neutrino lighter than $\alpha$ neutrino')
to generate a large asymmetry. The instability is however not enough.
It can be demonstrated that for mixing angles
$\theta_0\gg 10^{-2}$ incoherent scatterings of neutrinos do not destroy the
coherent evolution of the quantum state at  the resonance and the MSW effect
can occur as, for example, in  the Sun enterior. In this situation 
an asymmetry generation has been reported but only to values $10^{-7}-10^{-6}$ 
\cite{Kimmo}. The situation is completely different for small mixing angles
$\theta_0\ll 10^{-2}$. In this case collisions break the coherent 
evolution of the neutrino state at the resonance and supress the MSW effect, 
if the total asymmetry is less than a {\em border} value 
$L_b\simeq 10^{-6}\,(10^{-2}/\sin 2\theta_0)^{2/3}\,(|\delta m^2|/{\rm eV}^2)^{1/3}$. 
Thus in the beginning one can speak of a {\em collision dominated regime} \cite{comment}. 
Naively one could think that in this situation mixing
is frozen and there is no push for the beginning of the unstable regime.
This is not only false but even it is true that during the collision dominated regime
the push, at small mixing angles, is more powerful than 
at large mixing angles  in a MSW dominated regime. What happens is that
the MSW effect is replaced by another effect that relies on the
decoherent effect of collisions who make the quantum state to collapse
in one of the two flavor eigenstates with an amplitude depending on the 
matter mixing angle \cite{fv}. The generation occurs first with an exponential rate 
but is afterward made slower by the dependence of $y_{\rm res}^{\bar{\nu}}$ 
and $y_{\rm res}^{\bar{\nu}}$ on $L^{(\alpha)}$, a non linear effect. 
During the collision dominated regime the kinetic equations 
obey the statistical rule for which the system evolves toward a balance between direct and inverse processes and, if the equilibrium could be reached, one can show that after an initial generation of neutrino asymmetry this would be again destroyed. This however does not happen because collisions stop at neutrino decoupling, much before that the equilibrium can be reached.
Thus the system would be frozen in a out-of-equilbrium situation with an asymmetry
much larger than the baryon asymmetry, but still very small compared to the total number of neutrinos and antineutrinos. However if during the collision dominated regime the 
asymmetry could grow above the
value $L_b$, the MSW dominated regime can start. The MSW effect is completely 
independent on collisions and does not bring to a statistical equilibrium, 
while it discloses the real quantum nature of mixing for which the 
two neutrino flavors are two aspects of the same entity. 
What happens when the MSW regime starts ? The collision dominated regime of
the asymmetry growth has the effect to make $y_{\rm res}^{\nu}\neq y_{\rm res}^{\bar{\nu}}$.
If we still consider a positive initial asymmetry, then in this case at the end of
the collision dominated regime one has $y_{\rm res}^{\nu}\gg 10$, that means neutrinos
are well in the tail of the distribution, while $y_{\rm res}^{\bar{\nu}}\sim 0.1-1$
according to the choice of mixing parameters. From this moment their
evolution is approximately described by the expression $y_{\rm res}^{\nu}\propto L/T^2$,
while $y_{\rm res}^{\bar{\nu}}\propto (LT)^{-4}$. This means that the 
neutrino resonant momentum will continue to increase and thus 
neutrino mixing will be completely switched off.
 On the other hand, when the asymmetry gets a value comparable to the number 
of resonant antineutrinos, the asymmetry can increase only if $y^{\bar{\nu}}_{\rm res}$ 
starts to increase again and thus the power of the asymmetry evolution, 
$-d\ln L/d \ln T$, has to be necessarily less than $4$. 
In this way the anti-neutrino resonant momentum will span all the distribution
and if the MSW occurs adiabatically all anti-neutrinos will be converted \cite{fv2}. 
Thus the asymmetry will continue to increase until $y_{\rm res}^{\bar{\nu}}\stackrel{>}{\sim}10$. 
The maximum obtainable value of the asymmetry 
corresponds to a situation in which all anti-neutrinos have been converted 
and this is equivalent to have $L_{\nu_{\alpha}}=3/8$. 
This means that the MSW regime has the effect to push 
the system even further from the equilibrium, 
`completing the job' started by the collision dominated regime.

\section{Borders of neutrino asymmetry generation}

With Robert Foot \cite{ropa} we have determined the borders of 
the generation of neutrino asymmetry at mixing angles
$\sin^2 2\theta_0\leq 10^{-7}$. A first necessary condition is that 
the collision dominated regime can bring the asymmetry to grow beyond the value 
$L_b$ in a way that the MSW dominated regime can start, as in a sort of `relay race'. 
We found numerically that this happens approximately for 
$\sin^2 2\theta_0\stackrel{>}{\sim} 2\times 10^{-10}\,(|\delta m^2|/{\rm eV}^2)^{1\over 8}$.
In {\bf figure 1} we show three examples of total asymmetry evolution for the
same value of $-\delta m^2/{\rm eV}^2=10^{-3}$ but for three 
different values of $\sin^2 2\theta_0=4\times 10^{-10}$, $10^{-9}$ and $10^{-8}$.
In the first case the condition for $|L^{(\alpha)}|$ to get larger than $L_b$ 
is not satisfied and the asymmetry generation stops below the value $L_b$, while 
in the other two cases it grows above and the generation can continue.
This high sensitivity to the mixing angle is due both to the fact that
conversion rates in the collision dominated regime are proportional to 
$\sin^2 2\theta_0$ \cite{fv} and to the fact that $L_b\propto (\sin 2\theta_0)^{-2/3}$. 
Thus when the mixing angle decreases, both the generation rate is weaker and 
also the level to be reached for the MSW regime to start is higher. 
In the other two examples, for $\sin^2\,2\theta_0 = 10^{-9}$ and $10^{-8}$, 
the border value is reached and the MSW dominated regime can start. 
The final value of the asymmetry is thus much larger. However in one case is much
lower than the adiabatic limit, while this is closely approached in the third example 
for $\sin^2\,2\theta_0 =10^{-8}$. How to explain this difference in the final 
values in the case that the MSW regime starts ? This can be explained studying 
the adiabaticity of the MSW effect and for this one has to calculate
the {\em adiabaticity parameter} at the resonance, 
defined as $\gamma_r\equiv |2\dot{\bar{\theta}}_m\,\bar{{\ell}}_m|_{\rm res}^{-1}$
where $\bar{\theta}_m$ and $\bar{{\ell}}_m$ are respectively the mixing angle and 
the oscillation length in matter for the anti-neutrinos. 
If $\gamma_r \gg 1$, then the MSW effect is fully adiabatic 
and all active anti-neutrinos will be converted into anti-sterile neutrinos. 
When $\gamma_r \stackrel{<}{\sim} 3$ not all anti-neutrinos are converted. 
This can be more quantitavely expressed by the Landau-Zener approximation 
for which the conversion efficiency is given by the factor $(1-\exp[-\pi\,\gamma_r/2])$. 
It is possible to calculate explicitly the adiabaticity parameter finding that 
$\gamma_r\simeq 10^{10}\,(\sin^2 2\theta_0\,L\,T)/|4+d\ln L/d\ln T|$. 
From this expression and from the fact that 
$y_{\rm res}^{\bar{\nu}}\propto (LT)^{-4}$
it is possible to show that the final value of the asymmetry
for $y_{\rm res}^{\bar{\nu}}\simeq 10$ and $\gamma_r={\rm const}$, 
will be described, in the space of mixing parameters, by curves 
$\sin^2 2\theta_0\,(\delta m^2)^{1/4}={\rm const}$. This is confirmed by the 
numerical calculations that are shown in {\bf figure 2}. It is evident how
for $\sin^2 2\theta_0\stackrel{>}{\sim} 10^{-9}$ 
the study of the adiabaticity of the MSW effect reproduces the
right iso-final asymmetry curves with the correct power law (even
the value of the constant is quite well reproduced, see \cite{ropa}). 
On the other hand, at small mixing angles, this power law turns into the one
originating from the condition $|L^{(\alpha)}|>L_b$ at the end of the collision dominated regime. 
In this analysis we did not study the asymmetry generation for $\sin^2 2\theta_0> 10^{-7}$.
In \cite{ropa2} we showed how an asymmetry is produced at larger
mixing angles but the phenomenon of rapid oscillations appears for 
$\sin^2 2\theta_0\stackrel{>}{\sim} 10^{-6}$, something still 
not fully understood. We can however conclude that the neutrino asymmetry generation
from active-sterile neutrino oscillation is a mechanism that is well understood
in all his basic features, even though further invetigations will be necessary 
at large mixing angles, also to complete the borders of the generation 
shown in figure 2.

\begin{figure}
\psfig{figure=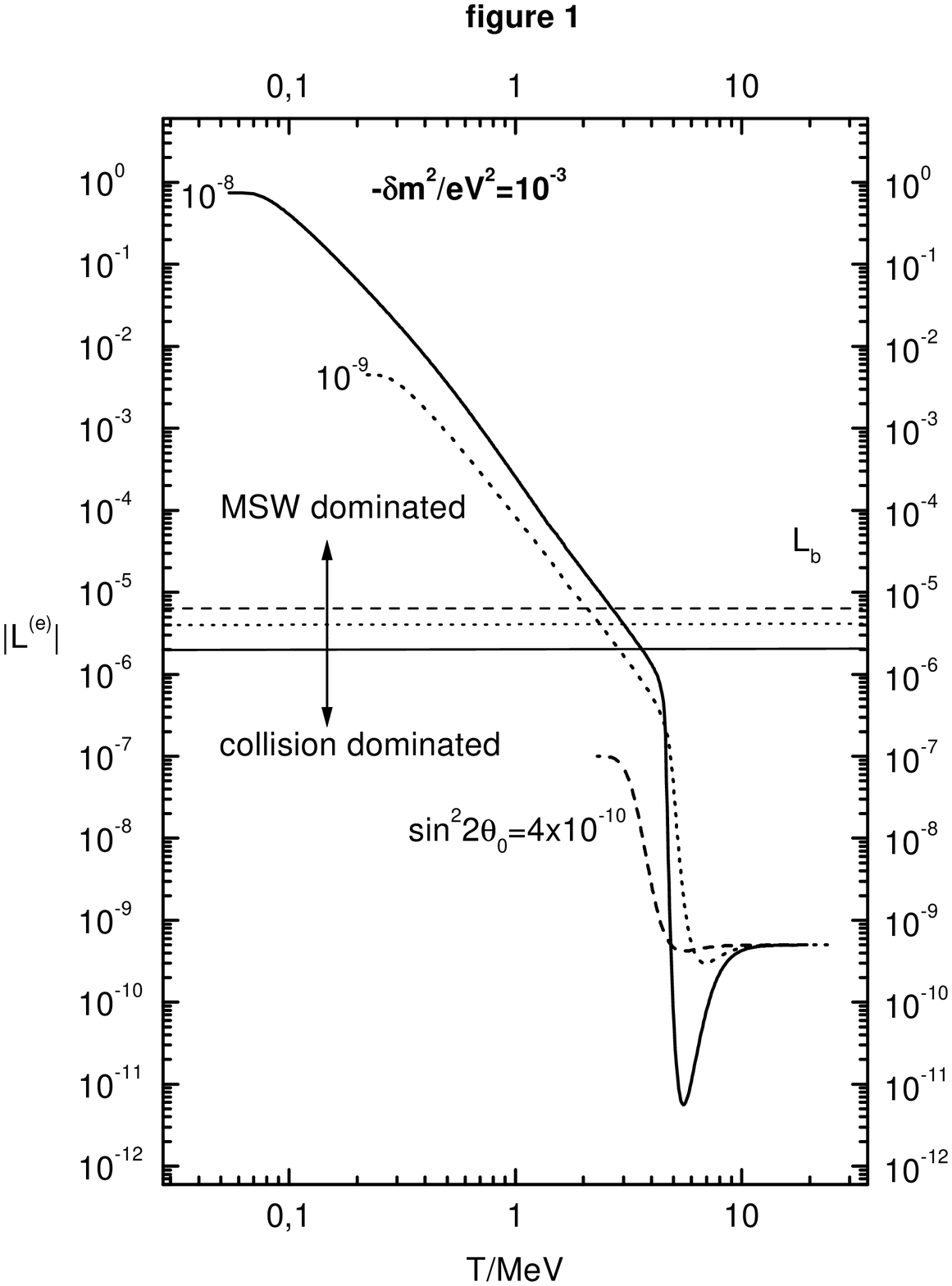,height=7cm,width=7cm}
\vspace{-10mm}
\hspace{-10mm}
\psfig{figure=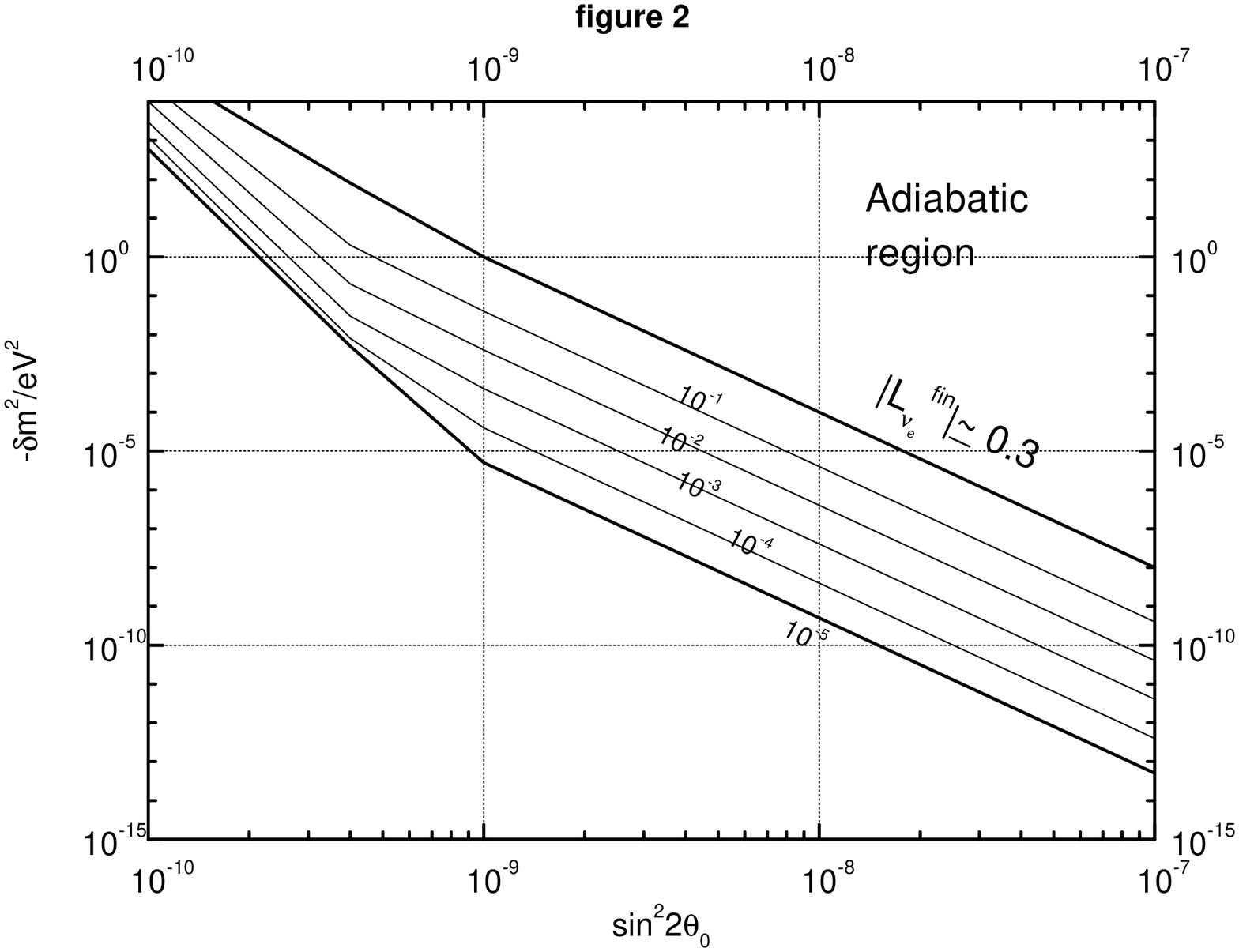,height=7cm,width=10cm}
\end{figure}
\section*{Acknowledgments}
The results on the borders of the generation have been obtained in
collaboration with Robert Foot \cite{ropa} and I am indebted to him 
for many discussions. PDB  wishes to acknowledge the warm hospitality 
of all people at the School of Physics  of Melbourne University during 
the period of his one year stay supported by an INFN research fellowship.

\end{document}